\documentclass[aps,prd,twocolumn,preprintnumbers,nofootinbib,superscriptaddress,amsmath]{revtex4-2}

    \usepackage{graphicx}
    \usepackage[caption=false]{subfig}
    \usepackage[usenames,dvipsnames]{xcolor} 
    \usepackage{pgfplots}
    \usepackage[utf8]{inputenc}
    \usepackage{color}
    \usepackage{hyperref}
    \usepackage[normalem]{ulem} 
    \usepackage{physics}
    \usepackage{enumitem}
    \usepackage{comment}
    \usepackage{bm}
    \usepackage{aas_macros}
    \usepackage[capitalise]{cleveref}
    \usepackage{appendix}
    \usepackage{multirow}
    \usepackage{lipsum}
    
\hypersetup{
  breaklinks = true,
  colorlinks   = true, 
  urlcolor     = blue, 
  linkcolor    = blue, 
  citecolor   = blue 
}
    \allowdisplaybreaks[1]

\usepackage{float}

\newcommand{\splitatcommas}[1]{%
  \begingroup
  \begingroup\lccode`~=`, \lowercase{\endgroup
    \edef~{\mathchar\the\mathcode`, \penalty0 \noexpand\hspace{0pt plus 1em}}%
  }\mathcode`,="8000 #1%
  \endgroup
}

\begin{document}
\title{A model-independent assessment of the late-time dark energy density evolution}
\author{Rayff de Souza}
\email{rayffsouza@on.br}
\affiliation{Observatório Nacional, Rio de Janeiro - RJ, 20921-400, Brazil}
\affiliation{School of Physics and Astronomy, University of Nottingham, University Park, Nottingham NG7 2RD, United Kingdom}

\author{Agripino Sousa-Neto }
\email{agripinoneto@on.br}
\affiliation{Observatório Nacional, Rio de Janeiro - RJ, 20921-400, Brazil}

\author{Javier E. González}
\email{javiergonzalezs@academico.ufs.br}
\affiliation{Universidade Federal de Sergipe, São Cristóvão - SE, 49107-230, Brazil}

\author{Jailson Alcaniz}
\email{alcaniz@on.br}
\affiliation{Observatório Nacional, Rio de Janeiro - RJ, 20921-400, Brazil}

\date{\today}

\begin{abstract}
Combined measurements of Baryon Acoustic Oscillations (BAO) from the Dark Energy Spectroscopic Survey (DESI), the Cosmic Microwave Background (CMB) and Type Ia Supernovae (SN Ia), have recently challenged the $\Lambda$-Cold Dark Matter ($\Lambda$CDM) paradigm, indicating potential evidence for a dynamical dark energy component. These results are usually obtained in the context of the dark energy equation-of-state (EoS) parameterizations, generally implying in phantom-crossing at intermediate redshifts. However, a general mapping between these parameterizations that yields approximately the same background observables clouds the inference of the true nature of dark energy in the context of these parametric methods. In this work, we propose a model-independent reconstruction of the dark energy density, which is more directly constrained than its EoS, based on the Gaussian Process (GP) regression method with the use of DESI DR2 BAO data and the Pantheon+, Union3 and DESY5 SN Ia samples. In addition, we perform a statistical comparison between the energy densities of $\Lambda$, a non-phantom thawing quintessence-type dark energy, and the Chevallier-Polarski-Linder parameterization with the reconstructed function. We find that all models agree with the GP reconstruction at 95\% C.L., with the largest discrepancy coming from $\Lambda$CDM with DESY5 at low redshifts. Even in this case, our findings suggest that it may be premature to claim statistically significant evidence for evolving or phantom dark energy with current DESI and SN Ia measurements.
\end{abstract}




\maketitle



\section{Introduction}\label{sec:1}

Recent measurements of Baryon Acoustic Oscillations (BAO) by the DESI Collaboration \cite{DESI:2025zgx}, in combination with Cosmic Microwave Background (CMB) and Type IA supernovae (SNeIA) data, have challenged the standard $\Lambda$-Cold Dark Matter ($\Lambda$CDM) cosmological model, indicating a possible evidence for a dynamical dark energy (DDE) component in the Universe. Such claims are mainly based on a $\gtrsim 3\sigma$ discrepancy between $\Lambda$CDM and the mean distribution of dynamical dark energy equation-of-state (EoS) parameterizations, such as the Chevallier-Polarski-Linder \cite{Chevallier:2000qy,Linder:2002et} (CPL), $w(a) = w_0 + w_a(1-a)$, where $w(a)$ is the time-dependent ratio between the dark energy pressure and its energy density, and $a$ is the cosmological scale factor.

One can wonder how much the evidence against a cosmological constant depends on assumptions about the DDE component, an issue discussed in recent literature \cite{Sousa-Neto:2025gpj,Rodrigues:2025tfg,Li:2025ops,Gonzalez-Fuentes:2025lei,Chaudhary:2025vzy,Blanco:2025vva,Zhang:2025bmk,Mukherjee:2025fkf,Cline:2025sbt,Gialamas:2025pwv,SanchezLopez:2025uzw,Wolf:2025acj,Fazzari:2025lzd,Capozziello:2025lor,Wang:2025xvi,Dhawan:2025mer,Mukherjee:2025fkf,GuptaChoudhury:2025uff,Yang:2025kgc,Yao:2025wlx,Yang:2024kdo,Yang:2025mws,Das:2023rvg}. Other interpretations of DESI's findings have also been considered, for example, in the form of anisotropic cosmic expansion \cite{Camarena:2025upt}, modified recombination \cite{Mirpoorian:2025rfp}, {orthogonal linear combinations of the distances measures \cite{Efstathiou:2025tie}} and a phantom mirage phenomenon \cite{Liu:2025bss,Caldwell:2025inn}. In the context of the standard assumption of a generalized DDE component, the common approach of proposing a dark energy EoS, such as the well-known CPL, Barboza-Alcaniz \cite{Barboza08} (BA), and Jassal-Bagla-Padmanabhan \cite{Jassal_2005} (JBP) parameterizations, has the advantage that these extended models are nested with $\Lambda$CDM, i.e., they recover a cosmological constant as the limiting case of some bridging parameters. Thus, one can statistically confront these parameterizations against current data and search for evidence in favor of the generalized model, quantified by how many standard deviations the $\Lambda$CDM limit is away from the mean distribution. However, the level of discrepancy obtained in this manner significantly depends on the choice of the DDE modeling, ranging from well within 1-2$\sigma$ to the 4-5$\sigma$ threshold \cite{deSouza:2025rhv,DESI:2025zgx}. In addition, the latter level of discrepancy is often obtained within the framework of  generic $w_0,w_a$ parameterizations, which seem to provide a better fit to the data but have little to no physical motivation \cite{Abreu:2025zng,Toomey:2025xyo}.

\begin{figure}[t]
\centering
\includegraphics[width=\linewidth]{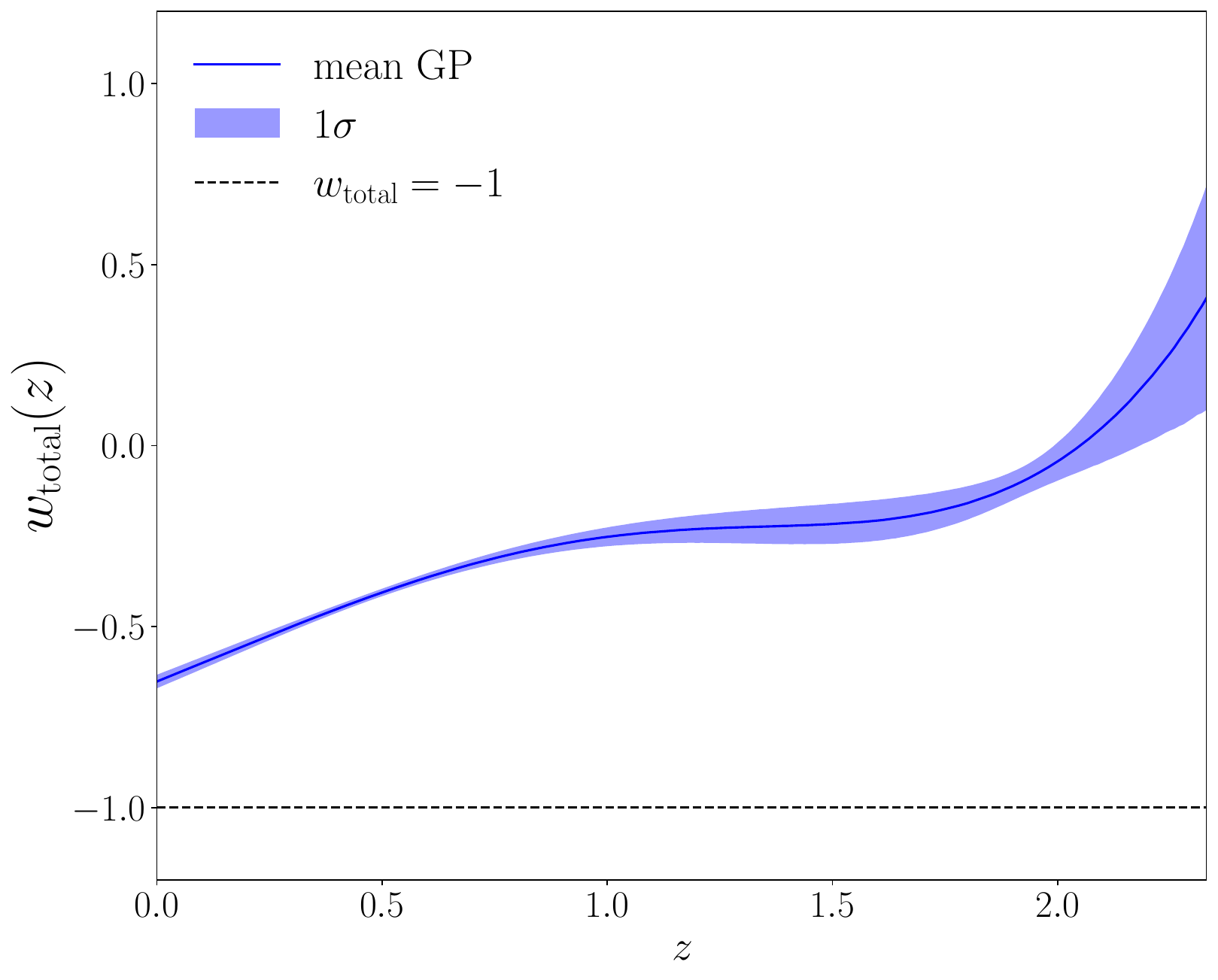}
\caption{Total equation of state of the Universe at low-redshifts, obtained via Gaussian Processes regression with DESI DR2 and Pantheon+ data.}
\label{w_total}
\end{figure}

\begin{figure*}[t]
\centering
\includegraphics[width=0.45\linewidth]{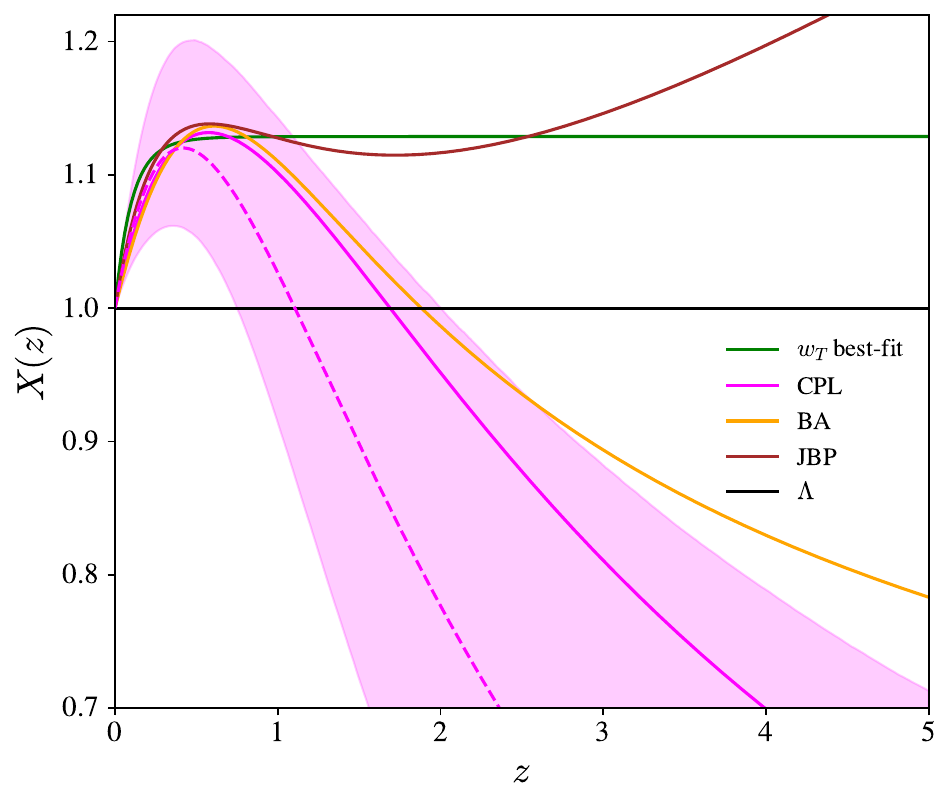}
\includegraphics[width=0.51\linewidth]{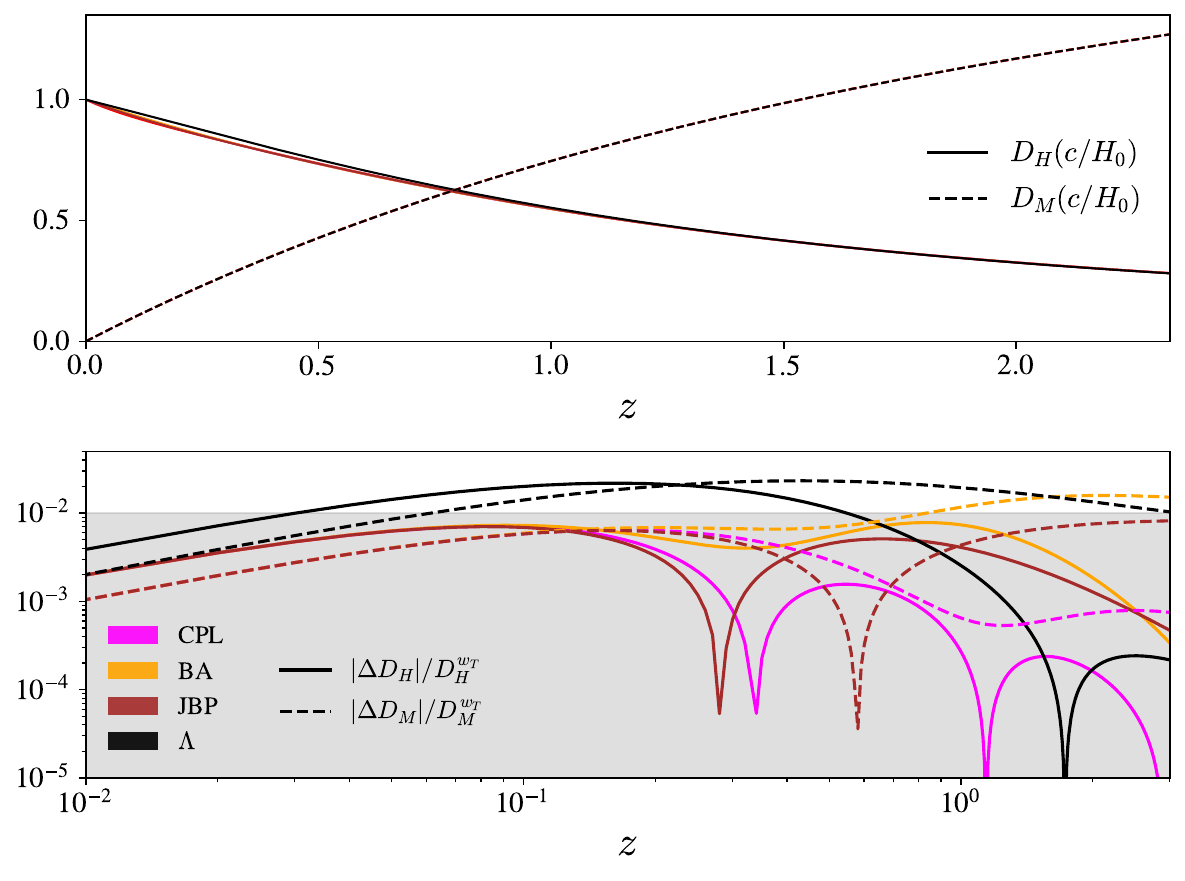}
\caption{\textit{Left}: Solid curves show the mapping between the dark energy densities for several common parameterizations, a cosmological constant and the $w_T(z)$ thawing quintessence. The dashed curve and the shaded region display the best-fit energy density and the 95\% C. L for CPL using the DESI BAO DR2+Planck 2018+DESY5 data combination, respectively. \textit{Right}: The upper panel shows the predictions for  $D_H(z)$ and $D_M(z)$ in units of $c/H_0$ for each of the solid curves depicted in the left plot, where all the curves, except for the $\Lambda$CDM prediction in black, are superimposed on top of each other. In the lower panel we highlight the relative difference between the models' distances and the best-fit prediction for the thawing quintessence parameterization. The grey shaded band denotes the region where the relative differences stay below the sub-percent threshold.}
\label{mapping}
\end{figure*}

However, the region in the $w_0 - w_a$ plane of these EoS parameterizations that is preferred by the data can be associated with the predictions of single-field scalar field models, due to a general mapping between their expansion rates that yields roughly the same background observables \cite{Wolf:2023uno,Wolf:2025jlc,Shlivko:2024llw}. In general, the combination of recent data requires that $\mathrm{d}w(a)/\mathrm{d}a|_{a = a_0} > 0$, implying a negative $w_a$, which is achieved by the thawing quintessence class of scalar field dark energy. Therefore, one may interpret the DESI results as initial evidence for a thawing quintessence component in the Universe. However, that same region of the $w_0 - w_a$ plane predicts (within the framework of $w_0,w_a$ parameterizations) that dark energy transitions to a period of phantom-like behavior in the past, in which $w < -1$ and the dark energy density grows with cosmic expansion. {This limit is forbidden for minimally coupled scalar-field models, as it implies in a Hamiltonian unbounded from below, which gives rise to negative energy states and ghosts \cite{Carroll:2003st,2006IJMPD..15.1753C,2004PhRvD..70d3543C}. 

{Moreover, a universe dominated by a phantom dark energy component tends to violate the Null Energy Condition (NEC), which ensures the geodesic focusing of light rays in curved spacetimes and is given by $T_{\mu\nu} k^\mu k^\nu \geq 0$, where $T_{\mu\nu}$ is the total energy-momentum tensor and $k$ is any null 4-vector \cite{Santos:2007bs,Kontou:2020bta,Rubakov:2014jja}. In a cosmological setting, the NEC translates into the condition $w_\text{total} \geq -1$. Thus, even if a single component, such as dark energy, has an EoS lower than -1, the NEC may not necessarily  be violated, as long as the ratio of the total pressure and the total energy density of the Universe stays above -1. In this context, since the DESI-suggested transition to phantom dark energy happens towards the past - when dark energy is increasingly sub-dominant - the NEC is not expected to be violated, as noted in \cite{Caldwell:2025inn}. 

In fact, we can relate the total EoS of the Universe to background observables in a simple manner via the Friedmann equations, such that $w_\text{total}(z) = -1 -\frac{3}{2}(1+z)\frac{D''_M(z)}{D'_M(z)}$, where $D_M$ is the cosmological transverse comoving distance and primes denote derivatives with respect to redshift. In Fig. \ref{w_total}, we show a Gaussian processes reconstruction of the total EoS with DESI BAO DR2 and Pantheon+ SN Ia data -- the details of the reconstruction method are given in Sec. \ref{sec:2}. Therefore, since the favored CPL-type dark energy is not phantom during the DE-dominated epoch, there is no NEC violation at low-redshifts, as would also be the case for a standard scalar field component.}

Nonetheless, because of their background ambiguity, current distance-based data is unlikely to differentiate between these parametric models, even if they predict entirely distinct behaviors for the dark energy component when extrapolated to redshift regions that are not currently probed by distance-based data.

To further illustrate this point, one can consider a simple thawing quintessence-inspired parameterization for the dark energy equation of state (EoS), such as \cite{Carvalho:2006fy,deSouza:2025rhv} $w_T(a) = -1 + \alpha a^{\beta}$, 
which is capable of mimicking the behavior of several scalar-field potentials. The green curve in the left panel in Fig. \ref{mapping} shows the resulting dark energy density evolution $X(z)$ for the best-fit $w_T(z)$ for the combination of DESI BAO DR2, Planck 2018 and DESY5 SNe data, i.e. $\alpha = 0.42$ and $\beta = 10.4$ \cite{deSouza:2025rhv}. As expected, the energy density remains constant throughout most of the cosmic expansion while the field is frozen, until its kinetic energy increases, making the field roll down its potential and bringing the energy density to its current value, in the usual thawing quintessence behavior. 

For comparison, we also show the evolution of the dark energy densities that better reproduce the expansion rate $H(z)$ of this $w_T$ configuration\footnote{We adopt the procedure outlined in \cite{2025JCAP...06..054S}, where we obtain the best-fitting pair $\{w_0,w_a\}$ by minimizing the relative difference $E_H = \sqrt{\frac{1}{z_\text{max}}\int_0^{z_\text{max}}\left(\frac{H_\text{M}(z|\{w_0,w_a\})-H_{w_T}(z,\alpha,\beta)}{H_{w_T}(z,\alpha,\beta)}\right)^2\mathrm{d}z}$, where $\text{M}$ stands for the specific model used in the mapping and we fix $z_\text{max} = 4$. Our results are not particularly sensitive to the choice of $z_\text{max}$ nor to the minimization function -- see also \cite{Shlivko:2024llw,Wolf:2023uno} for alternatives.}, assuming the CPL, BA and JBP parameterizations, as well as the constant-valued $\Lambda$CDM line. As can be seen, the qualitative behavior of the dark energy density is significantly different for all of those scenarios, showing phantom-like evolution in all cases but the cosmological constant and thawing quintessence. Nevertheless, except for $\Lambda$, they all have basically indistinguishable background observables, as represented by the superimposed red-toned curves of $D_H(z) = c/H(z)$ and $D_M(z) = c\int_0^z d z^\prime / H(z^\prime)$ in the upper right panel of Fig. \ref{mapping}. Although not identically equivalent, the relative difference between $D_H$ of the $w_0,w_a$ parameterizations and $w_T$ stays roughly at the sub-percent level (grey band in the lower right panel of Fig. \ref{mapping}), which makes it difficult to distinguish them with current background data. 

A cosmological constant, however, does not have the necessary degrees of freedom to well approximate the expansion rate predicted by a thawing quintessence dark energy component, which increases the relative difference between their observables, making them more easily distinguishable. In this sense, a departure from $\Lambda$ can be inferred in the context of quintessence-like and generic $w_0,w_a$ parameterizations, but discriminating between these generalized models is more intricate. Even so, it is clear that, although those models are seemingly degenerate at the level of the expansion rate, they are not degenerate at the level of the dark energy density.

Naturally, the resulting pair $\{w_0,w_a\}$ one obtains when performing this mapping procedure is not the same as the best-fit $\{w_0,w_a\}$ obtained through a statistical analysis. However, taking the CPL parameterization as a representative case, we also show in the left panel of Fig. \ref{mapping} the best-fit and the 95\% C.L region of the dark energy density for the same data combination, given by the dashed curve and the shaded band, respectively. Thus, the CPL-mapped curve in the $w_T$ parameterization reproduces the phantom-like qualitative behavior of the best-fit and remains well within the 2$\sigma$ region, illustrating the challenge of statistically distinguishing these models to a definitive degree.

In any case, one can compare the quality-of-fit of these parameterizations to infer the most probable model and, consequently, if the data really prefer a phantom behavior for dark energy. However, this approach has its own shortcomings. First, as discussed in \cite{Keeley:2025rlg}, although $\Lambda$CDM is easily nested within these parameterizations, they are generally not nested with each other, which does not allow for a clear visualization of the level of discrepancy between them from a frequentist point of view, in terms of standard deviation units. Moreover, a Bayesian evidence comparison, which could substitute the frequentist approach, is sensitive to the prior choices considered in the statistical analysis.

A related issue of assuming a EoS parameterization is that the dark energy density is the quantity that can be more directly constrained from distance-based data, since it is straightforwardly related to the cosmic expansion rate through the Friedmann equation. The dark energy EoS, on the other hand, is related to the expansion rate through an integral such that $H^2(z)/H_0 \propto X(z) \equiv \rho_{DE}(z)/\rho_{DE,0} = \exp\left\{ 3\int_0^z\frac{1+w_{DE}(z^\prime)}{1+z^\prime}dz^\prime \right\}$. In turn, $w_{DE}(z)$ is more loosely constrained, since it suffers from the smearing effect caused by the multiple integrals that relate it to cosmological distance measures \cite{Wang:2001ht}. Also, as noted in \cite{Wang:2004ru}, the inference of the EoS always requires one additional parameter than the energy density, since $w_{DE}(0)$ is a completely free parameter, whereas $X(0) = 1$. 

Therefore, based on the above considerations, embedding $\Lambda$CDM into a specific form of the dark energy equation of state could weaken the evidence for DDE and cloud the inference of the true nature of dark energy. This highlights the importance of extracting information about the dark energy density without relying on EoS parameterizations. In principle, a more direct data-driven inference of the dark energy density could guide us towards a deeper understanding of the dark energy behavior, be it phantom or not. To this end, we focus on inferring the dark energy density in a model-agnostic way based on the Gaussian Process (GP) regression method \cite{Seikel_2012}. This procedure has the advantage of expressing the dark energy density as a function of cosmological distance measures, which are directly reconstructed from available data on $D_H(z)$ and $D_M(z)$.

This work is organized as follows. In Sec. \ref{sec:2} we outline the methodology for reconstructing the dark energy density with the GP method, while the data combinations utilized in our analyses are discussed in Sec. \ref{sec:3}. In Sec. \ref{sec:4}, we present our results regarding the GP reconstruction and a statistical comparison of the thawing quintessence, CPL and $\Lambda$CDM models with the model-independent approach. We end this paper with a summary of the results and our main conclusions in Sec. \ref{sec:5}.

\section{Dark energy density reconstruction}\label{sec:2}

Assuming a negligible contribution from radiation at late-times, the Hubble parameter in a flat FLRW spacetime is written as
\begin{equation}\label{friedmann eq.}
    \frac{H^2(z)}{H_0^2} = \Omega_m (1+z)^3 + (1-\Omega_m) X(z)\; .
\end{equation}
where $H_0$ is the Hubble constant and $\Omega_m$ is the current matter density parameter. 

In order to perform an observation-based reconstruction, our goal is to use Eq.~(\ref{friedmann eq.}) to express $X(z)$ as a function of cosmological distance measures. To this end, we focus on the transverse comoving distance $D_M(z)$, since the distance-duality relates this quantity to the luminosity distance $D_L(z) = (1+z)D_M(z)$, which allows us to use both BAO and SNe data in our analysis. The former is obtained by inferring $D_M$ relative to the sound horizon at the drag epoch $r_d$, such that $\tilde{D}_M(z) \equiv D_M(z)/r_d$, given by
\begin{equation}\label{D_M}
    \tilde{D}_M(z) = \frac{c}{r_d}\int_0^z \frac{d z^\prime}{H(z^\prime)}\;\hspace{0.2cm} \text{ and }\hspace{0.2cm}\tilde{D}_M^\prime(z) = \frac{c}{r_d H(z)}\; .
\end{equation}
By evaluating the second equation above at the present time we can write $\tilde{D}_{M,0}^\prime \equiv \tilde{D}_M^\prime(0) = c/r_d H_0$. Substituting this expression and the expression for $\tilde{D}^\prime_M(z)$ into the left-hand side of Eq.~\eqref{friedmann eq.} and solving for the normalized dark energy density, we find:
\begin{equation}\label{X(z) as a func of D_M}
    X(z) = \frac{[\tilde{D}_{M,0}^\prime/\tilde{D}_M^\prime(z)]^2 - \Omega_m(1+z)^3}{1-\Omega_m}\;.
\end{equation}

When SN Ia data is included, $\tilde{D}_M(z)$ is related to the luminosity distance $D_L(z)$ via the distance-duality relation
, which also incorporates the absolute $M$ and apparent magnitude $m$, such that:
\begin{equation}\label{distance-duality}
    \tilde{D}_M(z) = \frac{D_L(z)}{(1+z)r_d} = \frac{10^{\frac{m - M - 25}{5}}}{(1 + z) r_d}\;.
\end{equation}
To transform $D_L$ into   $\tilde{D}_M$, we consider $r_d =147.09 \pm 0.26 \text{ Mpc}$ \cite{Planck2018-Cosmological_Parameters} (see Appendix \ref{appendixA}
for further methodological details). Therefore, to obtain the dark energy density through Eq.~\eqref{X(z) as a func of D_M}, it is necessary to reconstruct the  first derivative of $\tilde{D}_M(z)$, which is obtained by the GP regression of DESI BAO and SN Ia data. Moreover, there is also a dependence on $\Omega_m$, which can be treated as a random variable in the reconstruction, with distribution estimated from the CMB distance priors. The methods in which these observables are employed in the analysis are discussed in detail in Sec. \ref{sec:3}.


\subsection{Gaussian Processes}
To reconstruct  $\tilde{D'}_M(z)$ and $\tilde{D}_M(z)$ in a non-parametric way, we use a Gaussian Process regression, which is a collection of random variables, any finite number of which have a joint Gaussian distribution \cite{RasmussenW06}, that has already been explored in a variety of contexts in cosmology \cite{vonmarttens2025, Seikel:2012uu,Ghosh_2024,Gonzalez:2016lur,Gonzalez:2017fra,Bengaly:2020vly,vonMarttens:2018bvz,vonMarttens:2020apn}. Assuming that the observable data $y$ and the underlying function $f(z)$ are components of a single Gaussian Process, we can describe their joint behavior through the process itself. This process is fully specified by a prior mean function $\mu(z)$ and a covariance function (kernel) $k(z, z')$, which encode the expected value of the function and the correlation between values at different input points $z$ and $z'$, respectively.
For this work, we adopt the widely-used Squared Exponential Kernel, given by
\begin{equation}
\label{kernel}
    k(z, z') = \sigma^2 \exp\left(-\frac{(z - z')^2}{2l^2}\right)\;.
\end{equation}
This function depends on two hyperparameters: $\sigma$ and $l$, which controls the magnitude of the variation and determines the smoothness of the reconstructed function, respectively. In a widely used approximation, GP optimizes both hyperparameters by maximizing the log marginal likelihood of the dataset, whereas in the exact approach, it marginalizes over them \cite{RasmussenW06, Seikel_2012,Hwang:2022hla,vonmarttens2025}. In this work, we adopt the former approach and a zero-mean prior function to avoid biased results.



The kernel in Eq. \eqref{kernel} defines the covariance structure assumed in the Gaussian Process modeling of an observable, i.e., the $\tilde{D}_M-\tilde{D}_M$ data correlation at different redshifts. Since the derivative of a GP is itself a GP, it is possible not only to reconstruct the derivative of an observable but also to incorporate derivative data of the observable directly within a joint GP framework. In our case, the target reconstruction is $\tilde{D}_H(=\tilde{D'}_M)$, and we therefore combine both $\tilde{D}_M$ from DESI+SN Ia data and $\tilde{D}_H$ from DESI data. Consequently, it is necessary to compute the corresponding $\tilde{D}_M - \tilde{D}_H$ and $\tilde{D}_H - \tilde{D}_H$  covariance functions, given by 

\begin{equation} \label{gaussian_kernel01}
    \frac{\partial k(z,z')}{\partial z'} = \frac{(z - z')}{l^2} k(z,z')\; ,
 \end{equation}
and
\begin{equation} \label{gaussian_kernel11}
     \frac{\partial^2 k(z,z')}{\partial z \partial z'} = \frac{l^2-(z - z')^2}{l^4} k(z,z')\; ,
 \end{equation}
respectively.

The GP hyperparameters are estimated by maximizing the log marginal likelihood \footnote{ The quantities in bold represent matrices or vectors.}:

\begin{equation}\label{log_likelihood}
\begin{split}
    \ln {\mathcal{L}}=&-\frac{1}{2}
[\bm { D_{M},D_H}]^T [\bm {\tilde{K}}(\bm z,\bm z)+\bm {\tilde{C}}]^{-1}[\bm { D_{M},D_H}] \\
&-\frac{1}{2}\ln |\bm {\tilde{K}}(\bm z,\bm z)+\bm {\tilde{C}}|-\frac{n}{2}\ln 2\pi\;,
\end{split}
\end{equation}

\noindent where \( \bm{z} \) denotes the vector of all redshift measurements from the DESI+SN Ia datasets, while \( \bm{\tilde{K}}(\bm{z}, \bm{z}) \) represents the covariance matrix associated with the Gaussian Process modeling, \( \bm{\tilde{C}} \) corresponds to the covariance matrix of the total observational data, and \( n \) indicates the total number of data points in the sample. Explicitly, $\bm {\tilde{K}}$ can be written as

\begin{eqnarray}
    {\tilde{\bm K}}(\bm{z}, \bm{z})=\left[
    \begin{array}{cc}
        \bm K(\bm{z_{D_M}}, \bm{z_{D_M}}) & \bm K'(\bm{z_{D_M}}, \bm{z_{D_H}}) \\
         \bm K'^T(\bm{z_{D_M}}, \bm{z_{D_H}})& \bm K''(\bm{z_{D_H}}, \bm{z_{D_H}})
    \end{array}\right]\;,
\end{eqnarray}

\noindent being $\bm z_{D_M}$ and $\bm z_{D_H}$  the vectors of redshifts for  the  $\tilde{D}_M$ data and the $\tilde{D}_H$ data, respectively. The elements of these submatrices are computed using Eqs.~(\ref{kernel}),(\ref{gaussian_kernel01}) and (\ref{gaussian_kernel11}), following \( [\bm{K}(\bm{z}, \bm{z})]_{i,j} = k(z_i, z_j) \), \( [\bm{K'}(\bm{z}, \bm{z})]_{i,j} = \partial k(z_i, z_j)/\partial z_j \) and \( [\bm{K''}(\bm{z}, \bm{z})]_{i,j} = \partial^2k(z_i, z_j)/\partial z_i\partial z_j \). Similarly, the observational covariance matrix ${\tilde{C}}$ is given by

\begin{eqnarray}
    {\tilde{C}}=\left[
    \begin{array}{cc}
        \bm C_{\tilde{D}_M} & \bm C_{\tilde{D}_M,\tilde{D}_H} \\
      \bm   C_{\tilde{D}_M,\tilde{D}_H}& \bm C_{\tilde{D}_H}
    \end{array}\right]\; ,
\end{eqnarray}
where $\bm C_{\tilde{D}_M}$ denotes the covariance matrix for $\tilde{D}_M$ measurements, $\bm C_{\tilde{D}_H}$ for $\tilde{D}_H$, and $\bm C_{\tilde{D}_M,\tilde{D}_H} $ represents their cross-covariance. For the DESI DR2 data, all these $\bm C$-covariance matrices are diagonal. In contrast, for the DESI+SN Ia dataset,  $\bm C_{\tilde{D}_M}$  is constructed as a block diagonal matrix formed by the SN Ia and  BAO $\tilde{D}_M$ data blocks, $\bm C_{\tilde{D}_H}$ corresponds to the DESI $\tilde{D}_H$  data, and  $\bm C_{\tilde{D}_M,\tilde{D}_H}$ contains the cross-covariance of  the DESI $\tilde{D}_M-\tilde{D}_H$  data,  with zero entries for the SN Ia $\tilde{D}_M$  $-$ DESI $\tilde{D}_H$  correlations.

The Gaussian Process reconstruction of the mean function and the correlations between  values at different redshifts are calculated as described in Ref. \cite{RasmussenW06, Seikel_2012,Dinda:2024ktd,Gonzalez:2024qjs}. Specifically, the reconstruction of $\tilde{D}_H(z)(=\tilde{D}'_M(z))$ given the data and optimized hyperparameter values follows from the multivariate Gaussian posterior distribution, with the posterior mean given by

\begin{equation}
\begin{split}
\boldsymbol{\overline{\tilde{D}_H}}(\bm z_*)  = & \\
\begin{bmatrix}
\bm K'^T(\bm {z_{D_M}},\bm {z_*}), \bm K''(\bm {z_*},\bm {z_{D_H}})
\end{bmatrix} 
& \tilde{\bm A}^{-1} 
\begin{bmatrix} 
\tilde{\bm D}_M \\
\tilde{\bm D}_H 
\end{bmatrix}\;,
\end{split}
\end{equation}

\noindent and its associated covariance is given by

\begin{equation}
\label{CovDH}
\begin{split}
    \mathbf{Cov}(\tilde{\bm D}_H (\bm {z_*})) = \boldsymbol{K''}(\boldsymbol{z_*},\boldsymbol{z_*})  \\
- \begin{bmatrix}
\bm K'^T(\bm {z_{D_M}},\bm {z_*}), \bm K''(\bm {z_*},\bm {z_{D_H}})
\end{bmatrix}
\tilde{\bm A}^{-1} &
\begin{bmatrix}
\bm K'(\bm {z_{D_M}},\bm {z_*})\\ \bm K''(\bm {z_{D_H}},\bm {z_*}) 
\end{bmatrix}\;. 
\end{split}
\end{equation}

\noindent Here, $\tilde{\bm A}=\bm {\tilde{K}}(\bm z,\bm z)+\bm {\tilde{C}}$ and $\bm z_*$ corresponds to the vector of redshifts for the reconstruction targets.

In Fig. \ref{DM_DM_prime} we show the GP result for $\tilde{D}_M(z)$ and $ \tilde{D}_H(z) (=\tilde{D}_M^\prime(z))$, where the latter is used directly in the reconstruction of the dark energy density via Eq. \eqref{X(z) as a func of D_M}. We also display the data points and their associated error bars for each of the SN Ia samples and the second release of DESI BAO.

\begin{figure*}[t]
\centering
 \includegraphics[width=\linewidth]{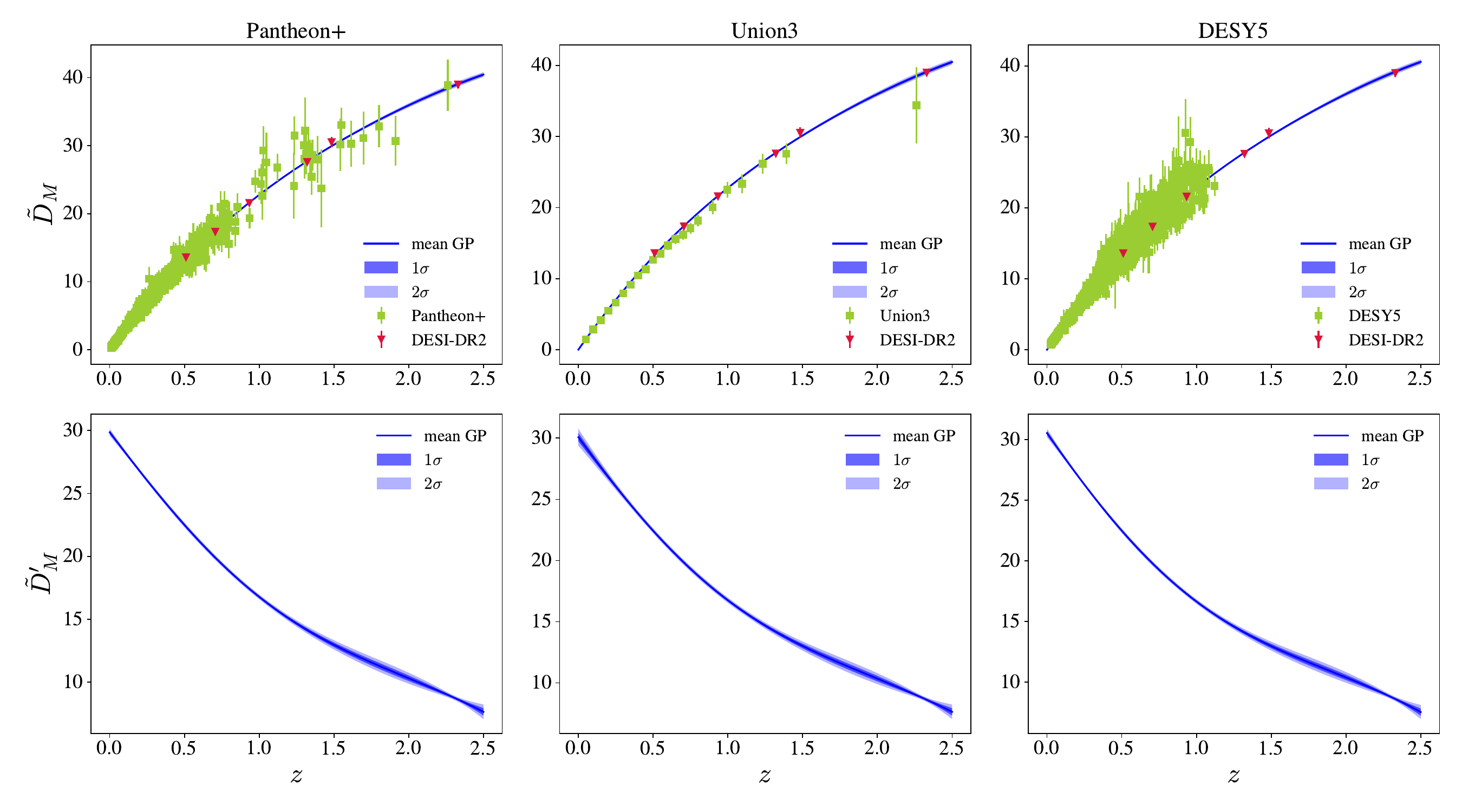}
 \caption{Gaussian Process reconstructions of $\tilde{D}_M$ (top) and $\tilde{D}^\prime_M (=\tilde{D}_H)$ (bottom) for Pantheon+, Union3, and DESY5 combined with DESI BAO.}
\label{DM_DM_prime}
\end{figure*}

\begin{table}[t]
\begin{ruledtabular}
\begin{tabular}{lcccc}
Tracer & $z_{eff}$&  $D_H/r_d$& $D_M/r_d$ & $r_{M,H}$
\\
\colrule
LRG1&  $0.510$ &  $21.863\pm0.425$ & $13.588\pm0.167$ &-0.459 \\
LRG2&   $0.706$ &  $19.455\pm0.330$ &$17.351\pm0.177$& -0.404\\
LRG3+ELG1& $0.934$ & $17.641\pm0.193$&$21.576\pm0.152$ &-0.416\\
ELG2  & $1.321$ &  $14.176\pm0.221$ & $27.601\pm0.318$ & -0.434\\
QSO & $1.484$ & $12.817\pm0.516$&$30.512\pm0.760$& -0.500\\
Ly$\alpha$& $2.330$ & $8.632\pm0.101$&$38.988\pm0.531$ & -0.431\\
\end{tabular}
\caption{\label{tab:table1}%
BAO measurements from the DESI (DR2) collaboration \cite{DESI:2025zgx}. The first and second columns show the tracers and the effective redshift, $z_{eff}$, while the third and forth columns show the $D_H/r_d$ and $D_M/r_d$ ratios, along with their $1\sigma$  limits. The last column presents the $D_M/r_d-D_H/r_d$ correlation.}
\end{ruledtabular}
\end{table}

\section{Observational Data}\label{sec:3}

This section provides a detailed account of the observational datasets employed in this work, encompassing BAO measurements from DESI DR2 and SNe data drawn from the Pantheon+, Dark Energy Survey Supernova Year-5 (DESY5), and Union3 compilations.

\subsection{DESI BAO}

The BAO measurements provided by DESI DR2 \cite{DESI:2025zgx} are reported in three forms: the three-dimensional, the transverse, and the radial modes. Consistent with the formulation introduced in the previous section, our analysis focuses on the radial mode, $D_H/r_d$, and transverse mode, $D_M/r_d$. Table~\ref {tab:table1} presents the measurement values for each tracer, namely Luminous Red Galaxy (LRG), divided into two samples (LRG1, LRG2), the combined LRG3 and Emission Line Galaxy (ELG1), ELG2, Lyman-$\alpha$ forest, and QSO sample. 

\subsection{Type IA Supernovae}

\subsubsection{Pantheon+}

The Pantheon+ compilation integrates 20 distinct Type Ia supernova datasets, collectively spanning the redshift interval $0.00122 \leq z \leq 2.26137$ \cite{Scolnic_2022}. It comprises 1701 light curves associated with 1550 spectroscopically confirmed SNe~Ia. To mitigate the influence of peculiar velocity uncertainties, 111 events with $z < 0.01$ were excluded from the analysis. The final working sample thus consists of 1590 light curves, covering $0.01016 \leq z \leq 2.26137$. The corresponding data products, including the full covariance matrix, are publicly available via the Pantheon+ GitHub repository\footnote{\url{https://github.com/PantheonPlusSH0ES/DataRelease/tree/main/Pantheon\%2B_Data}}.

\subsubsection{Union3}

The Union3 compilation \citep{Rubin2025} is an up-to-date version of the Union series, consisting of
2087 spectroscopically classified type Ia supernovae collected
from 24 observational surveys spanning the past decades. Union3 adopts
an approach based on the hierarchical Bayesian framework \textsc{Unity1.5} \citep{Rubin2015},
which consistently accounts for selection effects, intrinsic scatter,
and systematic uncertainties, and uses the SALT3 light-curve fitter \cite{Kenworthy2021}. The data are presented in binned form, in which the 2087 supernovae are grouped into 22 redshift points covering the range $0.05 \leq z \leq 2.26$.

\subsubsection{DESY5}

The DESY5 data set encompasses 1635 supernovae observed by the Dark Energy Survey (DES) collaboration over the range $0.0596 \leq z \leq 1.12$, supplemented by 194 high-quality external SNe~Ia at $z < 0.1$, yielding a total of 1829 objects. Events with photometric magnitude uncertainties exceeding $\delta m > 1$ — unlikely to correspond to genuine SNe~Ia within the DES photometric classification — were removed, resulting in a refined sample of 1754 supernovae. The complete catalog and associated resources are accessible through the DES collaboration’s GitHub repository\footnote{\url{https://github.com/des-science/DES-SN5YR}}.

\subsection{Cosmic Microwave Background}

We use CMB data from the latest release of the Planck Collaboration \cite{Planck:2019nip} in the form of the CMB distance priors on the shift parameter $R$, the acoustic scale $l_A$ and the physical baryon density parameter $\omega_b$ \cite{Chen:2018dbv}. This approach effectively compresses the full CMB likelihood and is shown to provide precise constraints on late-time cosmology. The parameters $R$ and $l_A$ are defined by
\begin{equation}\label{CMB distance priors}
    R = \frac{\sqrt{\Omega_m H_0^2} D_M(z_*)}{c}\;,\hspace{0.5cm}l_A = \frac{\pi D_M(z_*)}{r_s(z_*)}\;,
\end{equation}
with $r_s(z)$ being the comoving sound horizon and $z_*$ the redshift of photon-decoupling.

Taking the ratio of the above equations and substituting $\tilde{D}_{M,0}^\prime = c/r_d H_0$ yields
\begin{equation}\label{Omega_m CMB}
    \Omega_m = \left[ \frac{\pi R \tilde{D}_{M,0}^\prime r_d}{l_A r_{s}(z_*)} \right]^2\;.
\end{equation}

As shown in \cite{Dinda:2024ktd, Dinda:2024kjf}, assuming standard pre-recombination physics, $r_d$ and $r_s(z_*)$ are functions of only $\omega_b$ and the total physical matter density parameter $\omega_m$. The latter, on the other hand, can be obtained through the CMB distance priors by noting that $R/l_A = (\sqrt{\omega_m}/3000 \pi)r_{s_*}(\omega_b,\omega_m)/\text{Mpc}$. More details on this derivation are given in Appendix \ref{appendixA}.

Therefore, Eq. (\ref{Omega_m CMB}) essentially estimates $\Omega_m(R, l_A, D'_{M,0}, \omega_b, \omega_m)$, where the quantities in parenthesis are known distributions of random variables. We extract the mean and standard deviation of $\tilde{D}'_{M,0}$ from the reconstructed $\tilde{D}'_{M}(z)$ and we use the means and covariance of $\{R, l_A,\omega_b,\omega_m\}$ as reported in \cite{Chen:2018dbv} -- see Appendix \ref{appendixA}. 

In this context, we perform a Monte Carlo (MC) sampling by assuming a multivariate normal distribution on $\Omega_m$, where we find a corresponding distribution for each SN Ia used in the GP reconstruction, which yields slightly different values of $\tilde{D}'_{M,0}$ -- see Table \ref{tab:table3}. Thus, we use those results for $\Omega_m$ as our CMB data when performing the reconstruction of $X(z)$ per Eq. (\ref{X(z) as a func of D_M})\footnote{A caveat is in order here: the distributions of the CMB distance priors are obtained by assuming $\Lambda$CDM as the fiducial model. However, the results of \cite{Chen:2018dbv} are stable to modifications of late-time cosmology - such as DDE scenarios - with negligible changes in the posterior distribution of the parameters.}, which has the advantage of avoiding an explicit assumption on the value of $H_0$.

\section{Results}\label{sec:4}


\begin{figure*}[t]
\centering
\includegraphics[width=\linewidth]{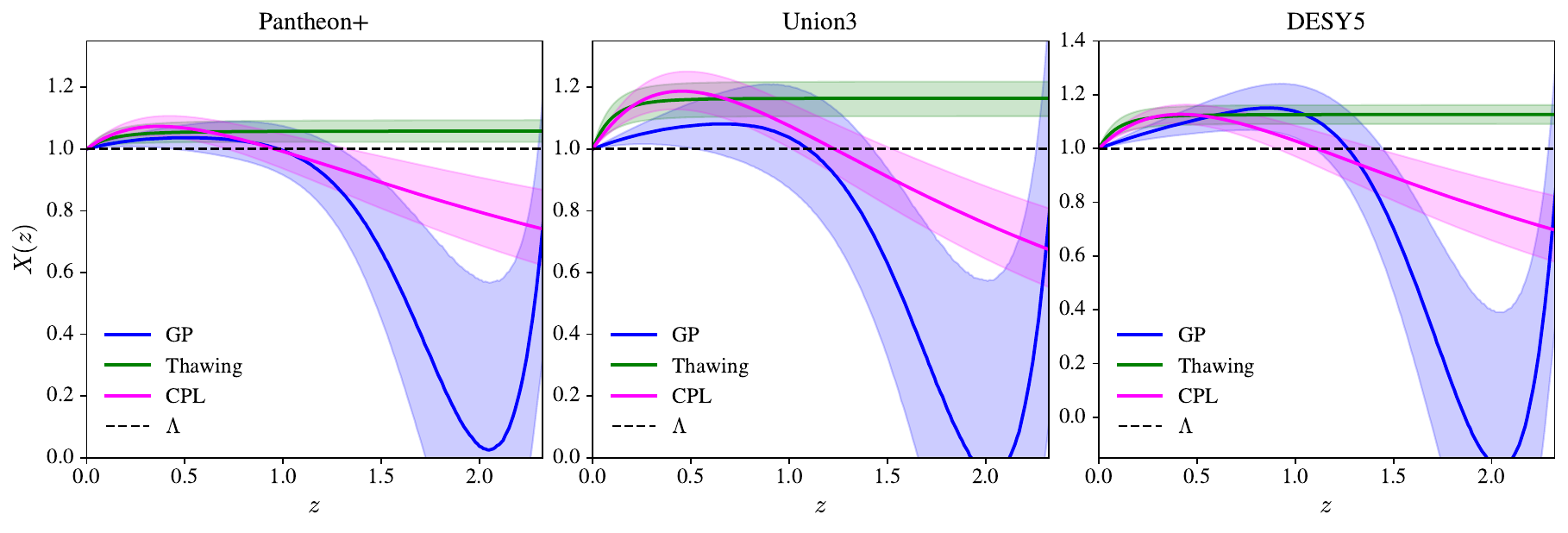}
\caption{GP dark energy density reconstruction and the statistical distribution of $X(z)$ for each parameterization. Each panel shows the results assuming the CMB+DESI DR2+SNe sample data combination. All shaded regions display the 68\% C.L.}
\label{X_z_all}
\end{figure*}

\begin{figure*}[t]
\centering
\includegraphics[width=\linewidth]{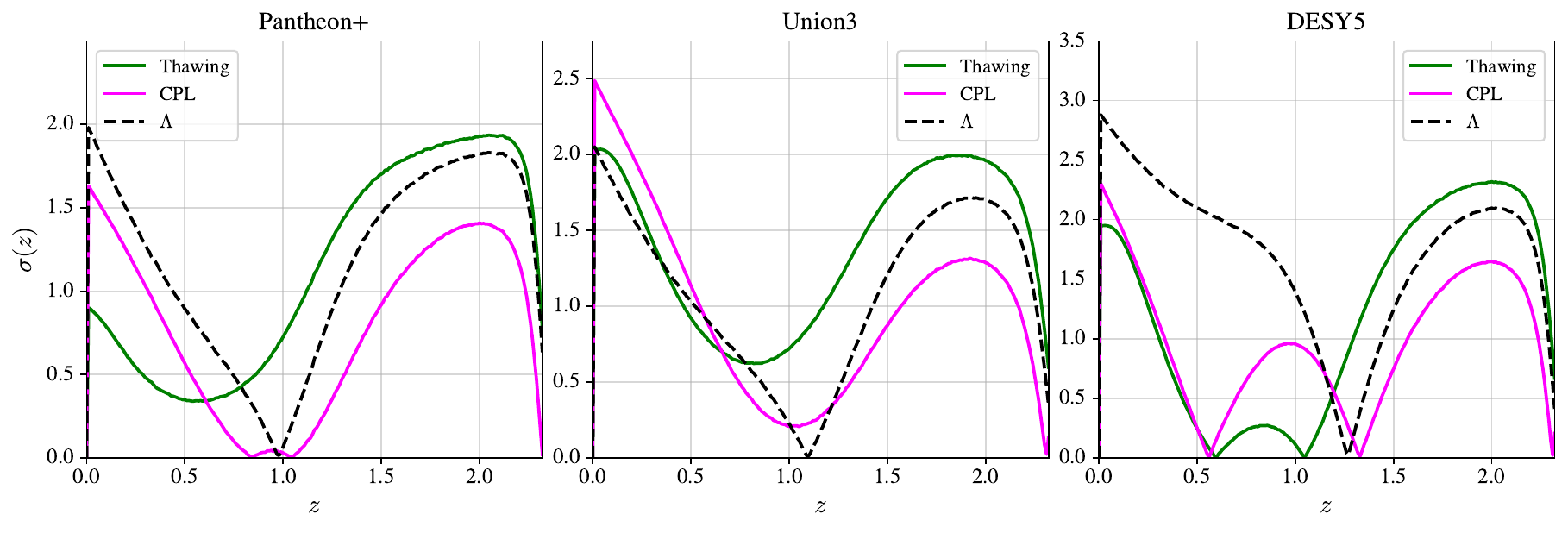}
\caption{Distribution of the standard deviation of each model relative to the GP reconstruction as a function of redshift, for each SNe sample used in the data combination.}
\label{sigma_z_all}
\end{figure*}

\begin{figure*}[t]
\centering
\includegraphics[width=\linewidth]{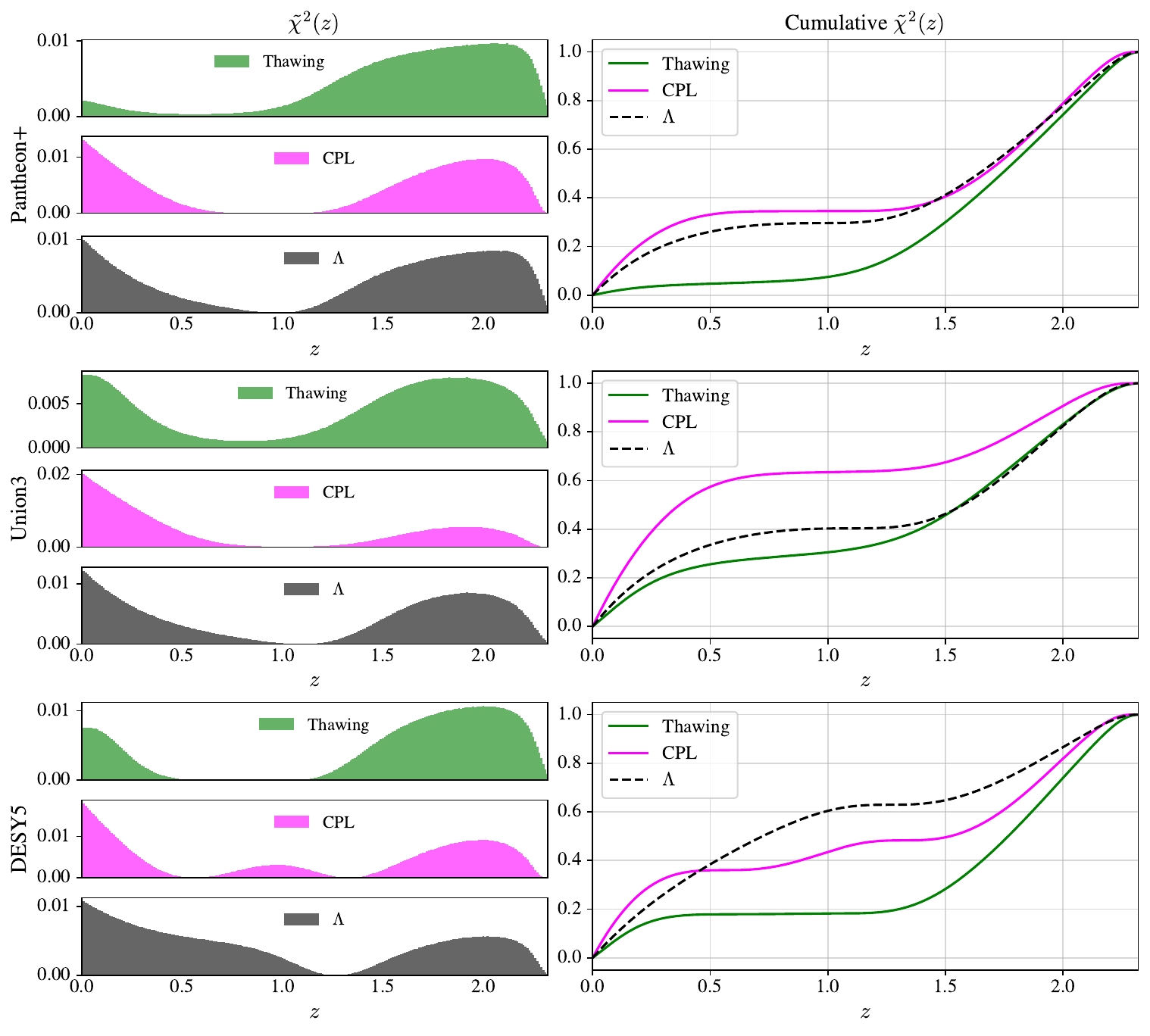}
\caption{\textit{Left}: Distribution of the fractional $\tilde{\chi}^2(z)$ measure relative to the GP reconstruction for each dark energy model. \textit{Right}: Cumulative distribution of the fractional $\tilde{\chi}^2(z)$ as a function of redshift. Each row corresponds to the results given a different SN Ia sample.}
\label{frac_chi2_and_cumulative_chi2_all}
\end{figure*}

Our main results are displayed in Fig. \ref{X_z_all}, where we show the GP reconstruction of the dark energy density for the DESI data combinations with the three SN Ia samples considered, Pantheon+, Union3 and DESY5 respectively. The dark energy density for each data combination is calculated from the GP reconstruction  of $\tilde{D'}_M(z)$ and the distribution of $\Omega_m$ via Eq. \eqref{X(z) as a func of D_M} by performing a Monte Carlo sampling accounting for the covariance between $\tilde{D'}_{M,0}$ and $\tilde{D'}_{M}(z)$ estimated in Eq. \eqref{CovDH}.  We also overlay the plots with the dark energy density distributions assuming the CPL and the thawing quintessence parameterizations -- obtained via Markhov Chain Monte Carlo (MCMC) sampling with the same datasets -- as well as the cosmological constant line. All curves include the 68\% C.L in the shaded region. 

As shown across the three panels in the blue curves, the GP reconstruction preserves its approximate shape regardless of the SN Ia catalog. For the Pantheon+ sample, the parameterized dark energy densities well approximate the reconstructed function for $z \lesssim 1$, while the Union3 and DESY5 compilations show a larger discrepancy already at those lower redshifts. For all the datasets, the GP reconstruction features a sharp drop after $z \sim 1$, but the position of the peak changes to some degree in each case, characteristic of a dark energy EoS that crosses the phantom barrier at slightly different redshifts. In addition, all GP curves return to larger values after $z \sim 2$, driven by the highest redshift Ly$\alpha$ DESI point. 

Given the model-independent data-driven dark energy density, one is then instructed to compare which of the given parameterizations better fit the reconstructed function. However, as discussed in Sec. \ref{sec:1}, this procedure is not unique, as the distinct models are not nested within each other nor to the reconstruction. Arguably, the simplest method would be to consider the GP-reconstructed function as the fiducial dark energy density and employ a goodness-of-fit approach to each of the parameterizations. This procedure has the advantage of taking an unbiased dark energy density as the fiducial model (which stems directly from the data), which can then be compared with the statistical distribution of each parameterization obtained with the same data.

In this context, each parameterization, as well as the GP reconstruction, has a mean distribution of $X(z)$ for each of the $N$ redshift points. At first, we can quantify the overall discrepancy between a given parameterization and the GP function as a summation over $N$, given by the normalized $\chi^2$ measure:
\begin{equation}\label{chi2}
    \chi^2_n = \frac{1}{N}\sum_{i=1}^N\frac{\Delta_X^2(z_i) }{\sigma^2_\text{GP} (z_i) +\sigma^2_\text{M} (z_i)}\;,
\end{equation}

where $\Delta_X (z_i)= \bar X_\text{M} (z_i)- \bar X_\text{GP} (z_i) $ is the difference between the mean value predicted by the model and the mean value of the GP reconstruction, and $\sigma^2_{M}(z_i)$ and $\sigma^2_{GP}(z_i) $ are their respective variances at each of the $i = 1,\dots,N$ redshift points. 

The above operation effectively captures the statistical significance of the deviation from the GP reconstruction taking into account the entire redshift range probed by data. Also, we can convert it to a global standard deviation measure by taking the square-root of \eqref{chi2}, such that $\bar\sigma = \sqrt{\chi^2_n}$. When performing the reconstruction, we fix $N = 250$, although the results are not sensitive to this choice. Moreover, in order to inspect the discrepancy as a function of redshift and to localize the regions of highest tension, we use the redshift-dependent $\chi^2$, such that $\chi^2(z) = \Delta_X^2(z)/(\sigma^2_\text{GP}(z)+\sigma^2_\text{M}(z))$. Accordingly, the redshift-dependent standard deviation is taken as $\sigma(z) = \Delta_X(z)/\sqrt{\sigma^2_\text{GP}(z)+\sigma^2_\text{M}(z)}$. 

We show the results for $\sigma(z)$ in Fig. \ref{sigma_z_all} for thawing quintessence, CPL and $\Lambda$CDM with each of the SN Ia samples, as well as $\bar\sigma$ and the maximum value of $\sigma(z)$ in Table \ref{tab:table2}. In all the panels, there is a clear distinction between the low- and high-redshift ranges, mostly dominated by SN Ia and DESI data, respectively. Also, a common trend is thawing quintessence being the most discrepant parameterization at higher $z$ ($z \gtrsim 1 - 1.5$), in line with the apparent requirement of phantom behavior at those redshifts, which is not achieved by this class of models. Within this region, the energy density associated with $\Lambda$ is smaller than that of thawing quintessence, resulting in a closer agreement with the GP reconstruction, while CPL approximates it the most due to the aforementioned phantom evolution. This pattern is expected, since in this range the reconstruction is dominated by DESI BAO data, which aligns our results with the ones obtained by the DESI Collaboration favoring a phantom-type dark energy behavior \cite{DESI:2025zgx, DESI:2025fii}. Nevertheless, in all three panels all parameterizations stay within $\sim 2\sigma$ of the GP function at these redshifts.

 

\begin{table}[t]
\begin{ruledtabular}
\begin{tabular}{lcc}
 model & $\bar\sigma$ & $\sigma_\text{max}$ \\
\colrule
\multicolumn{3}{c}{{Pantheon+}} \\
\colrule
Thawing & $1.29$ & $1.93$  \\
CPL & $0.93$  & $1.63$ \\
$\Lambda$CDM & $1.30$  & $1.98$\\
\colrule
\multicolumn{3}{c}{{Union3}} \\
\colrule
Thawing & $1.46$  & $2.03$\\
CPL & $1.15$ & $2.49$ \\
$\Lambda$CDM &$1.22$& $2.05$ \\
\colrule
\multicolumn{3}{c}{{DESY5}} \\
\colrule
Thawing & $1.47$  & $2.32$ \\
CPL & $1.13$ & $2.30$ \\
$\Lambda$CDM & $1.82$ & $2.88$\\

\end{tabular}
\caption{Values for the global discrepancy $\bar\sigma$ and maximum discrepancy $\sigma_\text{max}$ -- for each parameterization and SNe sample -- with respect to the GP reconstruction.}
\label{tab:table2}
\end{ruledtabular}
\end{table}



At lower redshifts, however, the results become more sensitive to the choice of the SN Ia sample, as each individual compilation provides somewhat different constraints on the late-time cosmic expansion rate, implying distinct requirements for the dominant dark energy density. For the Pantheon+ compilation, thawing quintessence provides the best fit to the data, since it predicts a decaying dark energy density that sits in intermediate values between $\Lambda$ and CPL, closer to the GP reconstruction -- see left panel in Fig. \ref{X_z_all} -- while CPL shows a smaller deviation than $\Lambda$ due to the larger errors around the mean. With Union3, both the thawing and CPL parameterizations exhibit a higher dark energy density (middle panel of Fig. \ref{X_z_all}), which then pushes them further away from the GP function, while the cosmological constant discrepancy stays roughly similar. On the other hand, the DESY5 results present the most distinctive signatures. First, as shown in the right panel of Fig. \ref{X_z_all}, the GP reconstruction features a stronger variation as a function of redshift, producing a dark energy density $\sim 20$ \% higher than a cosmological constant at $z \sim 1$. Consequently, this sample produces the highest discrepancy with $\Lambda$CDM, with the deviation from the GP regression approaching the 3$\sigma$ mark at very low redshifts -- see right panel of Fig. \ref{sigma_z_all}. That is also consistent with the evidence from DESI against $\Lambda$CDM with the inclusion of DESY5 SN Ia data, when it is nested within the CPL parameterization \cite{DESI:2025zgx}. 

Apart from analyzing how the different parameterizations compare between each other given the GP reconstruction for each dataset, it is complementary to see how the deviations in each parameterization are distributed across redshift. To this end, we compute the fractional $\chi^2$ measure, $\tilde{\chi}^2(z) \equiv \chi^2(z) / \sum_{i=1}^N \chi^2(z_i)$, in order to evaluate the redshift regions that accumulate the largest discrepancy. We also inspect the build-up of $\tilde{\chi}^2(z)$ as the redshift increases via its cumulative distribution. These metrics are shown in the left and right columns of Fig. \ref{frac_chi2_and_cumulative_chi2_all}, respectively. For thawing quintessence, virtually all $\chi^2$ is concentrated in the DESI-dominated region with the Pantheon+ data set, while being more spread across redshift with the other samples. The opposite occurs for CPL with Union3 data, where the highest build-up of the discrepancy appears with low-redshift data, and it is more evenly distributed with the other compilations. A cosmological constant, on the other hand, shows more peculiar traits across the SN Ia datasets. It accumulates a smaller or roughly equal deviation with Pantheon+ and Union3 between low-$z$ and high-$z$, and a very pronounced build-up at low-$z$ with DESY5, with approximately 60\% of the discrepancy showing in $z \lesssim 1$.

All of these results are summarized in Table \ref{tab:table2}. Although the deviations from the GP reconstruction vary with redshift given the different SN Ia catalogs, the overall measure $\bar{\sigma}$ stays below 2 in the standard deviation units across the board. The most striking discrepancies come with the DESY5 sample, with a cosmological constant, in particular, reaching closer to the $2\sigma$ threshold and $\sigma(z)$ approaching $3\sigma$ at very low redshifts. Even so, we find a consistently lower discrepancy with $\Lambda$CDM with this dark-energy density reconstruction compared to EoS parameterization analyses, in line with the results obtained in \cite{Wang:2025vfb,Sousa-Neto:2025gpj}.

\section{Summary and Conclusions}\label{sec:5}
Previous analyzes performed by the DESI collaboration have shown evidence for a dynamical dark energy component in the Universe. These claims are mainly based on the discrepancy between the $\Lambda$CDM limit of generic $w_0w_a$ parameterizations and the mean distribution of these models, while its statistical significance varies between the different choices of data combination. As argued in Sec. \ref{sec:1}, although these results shed light on possible deviations from $\Lambda$CDM, they do not provide a direct comparison between competing models in a unique manner, as different parameterizations are not nested within each other and can be easily mapped into one another by matching their cosmic expansion rates, yielding approximately equivalent distance-based observables.

In this work, we proposed an investigation of the dark energy density function -- which is more directly constrained by observations than its EoS -- in a model-independent data-driven framework, without relying on a specific parameterization as the fiducial model. This procedure was carried out with the use of the Gaussian Process regression method, which is able to reconstruct cosmic distance measures based on its available data across different redshifts. Specifically, we used luminosity-distance SN Ia data (which can be converted to the transverse comoving distance using the distance-duality relation) by the Pantheon+, Union3 and DESY5 data, and the transverse and radial BAO distances from the second data release of the DESI Collaboration. That allows us to take the GP non-parametric function as the dark energy density behavior that is directly imposed by the data, and then compute the deviations of each parameterization from it in a consistent fashion. Also, by directly extracting the dark energy density from data non-parametrically, we aim to provide a more informative description of the dark energy density in the redshift range covered by the data, which would assist in breaking the background ambiguity between models that predict completely different qualitative behaviors for the dark energy component -- see Fig. \ref{mapping}.

Our results regarding the GP dark energy density reconstruction and the predictions for each parameterization are shown in Fig. \ref{X_z_all}, and the comparison between the parameterizations and the GP function is summarized in Figures \ref{sigma_z_all}, \ref{frac_chi2_and_cumulative_chi2_all} and Table \ref{tab:table2}. In summary, in terms of the global deviation across the entire redshift range, $\Lambda$CDM, CPL and thawing quintessence are all consistent with the GP reconstruction, with the overall deviation staying below 2$\sigma$. 

However, if the data are divided into low- ($z \lesssim 1$) and high- ($z \gtrsim1$) redshift regions, the different parameterizations deviate from the GP regression at distinct levels. In general, the thawing quintessence parameterization provides the best fit to the GP at low, SN Ia dominated redshift, while being less consistent at the DESI-dominanted region. The opposite happens for CPL, which is consistently the best parameterization at larger redshifts, in line with the phantom behavior of the dark energy EoS at $z \gtrsim 1$ implied by DESI. $\Lambda$CDM, however, shows a larger discrepancy with the DESY5 catalog, where the largest portion of its deviation from the GP comes from low redshift SN Ia data. This behavior is in agreement with the strongest claims in favor of the CPL parameterization from DESI BAO, which often comes when DESY5 data are included as an additional dataset \cite{DESI:2025zgx} -- see also \cite{Ong:2025utx} for a critique on the inclusion of DESY5 data in the context of parametric dynamical dark energy statistical inference.}

Even though DESI BAO data achieve percent level accuracy, the scarce number of data points at $z \gtrsim 1.5$ results in somewhat large errors in the GP reconstruction, which hinders this method's capability of definitively pointing out the best-fitting dark energy density profile, irrespective of the chosen SN Ia catalog. Therefore, even though the overall shape of the GP reconstruction imply phantom-crossing in the DESI probed redshift region, neither a cosmological constant nor a thawing-type quintessence component can be ruled out.

Therefore, we expect that with increasingly more accurate SN Ia and BAO observations, as well as a larger volume of observed tracers, these model-independent methods will become an increasingly sharper tool in the determination of the late-time dark energy behavior. In this sense, these analyses provide an agnostic approach to extracting information from the data, which becomes paramount to infer the true nature of dark energy in this precision era of cosmology.

\section*{Acknowledgements}
The authors thank Edmund Copeland for critical comments on the manuscript and Carlos Bengaly Jr. for valuable discussions. RdS and ASN are supported by the Coordena\c{c}\~ao de Aperfei\c{c}oamento de Pessoal de N\'ivel Superior (CAPES). JA is supported by Conselho Nacional de Desenvolvimento Científico e Tecnológico (CNPq) grant No. 307683/2022-2 and Funda\c{c}\~ao de Amparo \`a Pesquisa do Estado do Rio de Janeiro (FAPERJ) grant No. 299312 (2023). The development of this work was aided by the National Observatory Data Center (CPDON).

\appendix
\section{Matter density parameter from CMB distance priors}\label{appendixA}

At sufficient early times, the Hubble expansion is driven primarily by the energy densities of radiation and matter, where sound waves propagate in a tightly-coupled photon-baryon fluid. In this regime, with fixed CMB temperature, the sound horizon at early times can be written entirely in terms of the physical baryonic and total matter densities, $\omega_b$ and $\omega_m$, respectively, and is approximately given by \cite{Dinda:2024kjf,Dinda:2024ktd}:
\begin{widetext}
\begin{equation}\label{sound horizon CMB}
r_s (z) = A ~ {\log} \left\{ \frac{{2}+(1+z)(B +C)+2\big[(1+z)^2BC+ (1+z)(B+C)+1\big]^{1/2}}{(1+z)\big[{B+C}+2({B C})^{1/2}\big]} \right\}\; ,
\end{equation}
\end{widetext}
where
\begin{eqnarray}
    A = \frac{100}{\sqrt{105}}\frac{(T_\text{CMB}/2.7~\text{K})^2}{\sqrt{\omega_b \omega_m}}~\text{Mpc}\;,\\
    B = \frac{1}{31500}\frac{(T_\text{CMB}/2.7~\text{K})^4}{\omega_b}\;,\\
    C = \frac{1}{1+25000 \omega_m (2.7~\text{K}/T_\text{CMB})^4}\; ,
\end{eqnarray}
and we take $T_\text{CMB} = 2.7255~\text{K}$. 

The redshift of photon-decoupling is also expressed as:
\begin{equation}\label{z_photon_decoupling}
    z_* = 1048 \left( 1+0.00124 \omega_b^{-0.738} \right) \big( 1+g_1 \omega_m^{g_2} \big) ,
\end{equation}
where
\begin{eqnarray}
    g_1 = \frac{ 0.0783\, \omega_b^{-0.238} }{ 1+39.5\, \omega_b^{0.763}}\;,\hspace{0.3cm}g_2 &=& \frac{0.560}{ 1+21.1\, \omega_b^{1.81} }\; .
\end{eqnarray}

\begin{table}[t]
\begin{ruledtabular}
\begin{tabular}{lccccc}
 param & mean $\pm~1\sigma$ & $R$ & $l_A$ & $\omega_b$ & $\omega_m$ \\
\colrule
$R$ & $1.7502 \pm 0.0046$ & $1.0$ & $0.46$ & $-0.66$ & $0.97$ \\
$l_A$ & $301.471^{+0.089}_{-0.090}$  & $0.46$ & 1.0 & $-0.33$ & $0.34$ \\
$\omega_b$ & $0.02236 \pm 0.00015 $  & $-0.66$ & $-0.33$ & 1.0 & $-0.50$ \\
$\omega_m$ & $0.1460 \pm 0.0013$  & $0.97$ & $0.34$ & $-0.50$ & 1.0 \\
\hline
\multirow{3}{4em}{$\tilde{D}^\prime_{M,0}$} & $29.86 \pm 0.14$ & \multicolumn{4}{c}{{Pantheon+}} \\
& $30.07\pm 0.34$ & \multicolumn{4}{c}{{Union3}} \\
& $30.54\pm 0.18$ & \multicolumn{4}{c}{{DESY5}} \\
\hline
\multirow{3}{4em}{$\Omega_m$} & $0.3070 \pm 0.0033$ & \multicolumn{4}{c}{{Pantheon+}} \\
& $0.3114 \pm 0.0072$ & \multicolumn{4}{c}{{Union3}} \\
& $0.3211 \pm 0.0042$ & \multicolumn{4}{c}{{DESY5}} \\



\end{tabular}
\caption{Distribution of the CMB distance priors used in our analyses \cite{Chen:2018dbv}. For the correlated parameters, namely $R$, $l_A$, $\omega_b$ and $\omega_m$, we report in the last 4 columns the values of $r_{i,j}$, meaning the normalized covariance between parameters $i$ and $j$. In the last two multi rows we list the distribution of $\tilde{D}^\prime_{M,0}$ and the resulting $\Omega_m$ for each SN Ia sample.}
\label{tab:table3}
\end{ruledtabular}
\end{table}

Therefore, substituting \eqref{z_photon_decoupling} into \eqref{sound horizon CMB}, we obtain $r_{s,*} \equiv r_s(z_*)$ which is a function of only $\omega_b$ and $\omega_m$. 

Taking the ratio of Eqs. \eqref{CMB distance priors}, we find that:
\begin{equation}\label{ratio CMB dp}
    \frac{R}{l_A} = \frac{\sqrt{\omega_m}}{3000 \pi}\frac{r_s(z_*)}{\text{Mpc}}\; ,
\end{equation}
where we have substituted $H_0 = 100~h~\text{km}~\text{s}^{-1}~\text{Mpc}^{-1}$, $c = 3\times 10^5~\text{km}~\text{s}^{-1}$ and $\omega_m = \Omega_m h^2$ as the physical matter density parameter. In this case, we use $r_{s,*}$ in the form of \eqref{sound horizon CMB} -- \eqref{z_photon_decoupling} together with \eqref{ratio CMB dp}, to obtain $\omega_m$ by solving
\begin{equation}\label{omega_m from CMB}
f(\omega_m,R,l_A,\omega_b) = \frac{r_{s,*}(\omega_b,\omega_m)}{\text{Mpc}} \sqrt{\omega_m} - 3000\pi \frac{R}{l_A} = 0\; , 
\end{equation}
which yields $\omega_m(R,l_A,\omega_b)$. 

Thus, we perform a Monte Carlo sampling of $\omega_m$ by solving \eqref{omega_m from CMB} with the distributions of the CMB distance priors as given in \cite{Chen:2018dbv}. The resulting distribution of $\omega_m$, as well as its normalized covariances, are listed in Table \ref{tab:table3}.

For the sound horizon at the baryon drag epoch, we use the fitting formula in order to express $r_d(\omega_b,\omega_m)$, namely \cite{Brieden:2022heh,DESI:2025fii}
\begin{equation}\label{rd_fitting}
    \frac{r_d}{{\rm Mpc}} = 147.05 \left(\frac{\omega_{\rm m}}{0.1432}\right)^{-0.23} \left(\frac{\omega_{\rm b}}{0.02236}\right)^{-0.13}\;,
\end{equation}
where we have assumed the standard value $N_\text{eff} = 3.04$ for the effective number of relativistic species. The value of $r_d$ which we find using \ref{rd_fitting} with the distributions of $\omega_b,\omega_m$ given in Table \ref{tab:table3} is entirely consistent with the Planck TT+TE+EE+lowE+lensing result of $r_d(\text{Mpc}) = 147.09 \pm 0.26$ which we used in the distance-duality relation -- see Sec. \ref{sec:2}. 

Finally, we perform another MC simulation in order to obtain $\Omega_m(R,l_A,\omega_b,\omega_m,\tilde{D}^\prime_{M,0})$ as given by Eq. \eqref{Omega_m CMB}, where we express both $r_d$ and $r_{s,*}$ as functions of $\omega_b$ and $\omega_m$, according to Eqs. \eqref{rd_fitting} and \eqref{sound horizon CMB} -- \eqref{z_photon_decoupling}, respectively. We use the distribution of the parameters and its covariances as listed in Table \ref{tab:table3}, where each $\tilde{D}^\prime_{M,0}$ is taken directly from the GP reconstruction. We also show the resulting $\Omega_m$ for each SN Ia sample. As mentioned in the text, this approach effectively allows us to use information from the CMB in our analyses and bypasses an explicit assumption on the value of $H_0$.

\bibliography{bibliography}

\label{lastpage}

\end{document}